\documentclass[aps,prx,reprint,superscriptaddress,showpacs]{revtex4-1}
\usepackage{hyperref}
\usepackage{amsmath}
\usepackage{graphicx}
\usepackage{subfigure}
\usepackage{color}
\usepackage{bm}

\newcommand{\bra}[1]{\left\langle#1\right|}
\newcommand{\ket}[1]{\left|#1\right\rangle}
\newcommand{\plaq}[1]{\vcenter{\hbox{\includegraphics[height=8pt]{plaq#1}}}}
\newcommand{\plaqh}[1]{\vcenter{\hbox{\includegraphics[height=16pt]{plaq#1}}}}

\begin{document}
\title{Double-semion topological order from exactly solvable quantum dimer models}
\author{Yang Qi}\email{qiyang@tsinghua.edu.cn}
\affiliation{Institute for Advanced Study, Tsinghua University, Beijing 100084, China}
\author{Zheng-Cheng Gu}\email{zgu@perimeterinstitute.ca}
\affiliation{Perimeter Institute for Theoretical Physics, Waterloo,
  Ontario, Canada N2L 2Y5}
\author{Hong Yao}\email{yaohong@tsinghua.edu.cn}
\affiliation{Institute for Advanced Study, Tsinghua University, Beijing 100084, China}
\affiliation{Collaborative Innovation Center of Quantum Matter,
  Beijing, 100084, China}

\begin{abstract}
  We construct a generalized quantum dimer model on two-dimensional
  nonbipartite lattices including the triangular lattice, the star
  lattice and the kagome lattice. At the Rokhsar-Kivelson (RK) point,
  we obtain its exact ground states that are shown to be a fully
  gapped quantum spin liquid with the double-semion topological
  order. The ground-state wave function of such a model at the RK
  point is a superposition of dimer configurations with a nonlocal
  sign structure determined by counting the number of loops in the
  transition graph. We explicitly demonstrate the double-semion
  topological order in the ground states by showing the semionic
  statistics of monomer excitations. We also discuss possible
  implications of such double-semion resonating valence bond states to
  candidate quantum spin-liquid systems discovered experimentally and
  numerically in the past few years.
\end{abstract}

\maketitle

\section{Introduction}

Topological order~\cite{krs,Wen1989,Wen1991a} is a key concept with
increasing importance in studies for quantum many-body systems,
especially for strongly correlated electronic systems. Quantum states
with the so-called intrinsic topological order can be characterized by
patterns of long-range quantum
entanglement~\cite{LevinWen2006,PriskillKitaev2006,ChenPRB2010}, and
feature fractionalized excitations carrying fractional charge and
anyonic statistics. Besides being interesting on their own, quantum
states with certain topological orders may have interesting
applications for fault-tolerant quantum
computations~\cite{kitaev,DasSarmaRMP2006}. The experimentally
observed fractional quantum Hall (FQH) states~\cite{FQHE,Laughlin1983}
are seminal examples of the emergence of topological order in strongly
correlated electronic systems. The concept of topological orders has
been widely used in the studies of quantum spin
liquids~\cite{wenSL1,wenSL2,wenSL3} and high-$T_c$
superconductors~\cite{Anderson1987,krs,Senthil2000,LNW2006Review}.

Quantum dimer models~\cite{Rokhsar1988} were first introduced to study
quantum spin liquids and high-$T_c$ cuprate in the context of
short-range resonating valence bond (RVB)
states~\cite{Anderson1973,Anderson1987}. Later it was shown that on a
nonbipartite lattice, the ground state of quantum dimer models
described by the Rokhsar-Kivelson (RK) Hamiltonian~\cite{sondhi2001}
can realize a particular topological order---the Z$_2$ topological
order~\cite{wenSL2,subir1991}, which is among the simplest intrinsic
topological orders and it is also known as the toric-code topological
order~\cite{kitaev}. In recent years, more topological orders have
been realized systematically in exactly solvable models, {\it e.g.},
the string-net models~\cite{Levin2005} and Kitaev
models~\cite{Kitaev2006}. In the past few years, evidences of gapped
quantum spin-liquid ground states were reported in numerical
simulations of frustrated spin SU(2)-symmetric models using the
methods of density matrix renormalization group (DMRG)~\cite{Yan2011,Depenbrock2012,Jiang2012a} and pseudofermion
functional renormalization group (PFFRG)~\cite{Suttner2014}, and in
experiments on real materials~\cite{yamashita}. Short-range RVB wave
functions (or, equivalently, dimer wave functions) with Z$_2$
topological order are often considered as candidate ground
states. Nonetheless, recent DMRG studies reported indirect evidence
that the Z$_2$ topological order might not describe the ground state
of the kagome spin-1/2 Heisenberg
model~\cite{HeShengChen2014,ShengSL,Bauer2014,He2014X,Barkeshli2014}.

Therefore, we naturally ask the following question: Can short-range
RVB-type wave functions in two-dimensional (2D) support a different
topological order than the Z$_2$ one? Our answer is positive by
constructing an exactly solvable Hamiltonian of quantum dimers (or
bond singlets) whose ground state has the double-semion topological
order~\cite{FNSWW04,Levin2005} instead of the Z$_2$ or toric-code
topological order. Our construction is quite simple as we only change
the phase of resonant terms in the original RK Hamiltonian, as follows:
\begin{equation}
\label{eq:iRK}
H=\sum_{P_\alpha}
\big[(it\ket{P_\alpha}\bra{P_{\tilde\alpha}}+\text{H.c.})
      +v(\ket{P_\alpha}\bra{P_\alpha}
        +\ket{P_{\tilde\alpha}}\!\bra{P_{\tilde\alpha}})\big],
\end{equation}
where $P_1=\plaq1$, $P_2=\plaqh5$, $P_3=\plaq4$, and
$P_{\tilde\alpha}$ is obtained by flipping the dimer configurations of
$P_\alpha$. Note that in the original RK
Hamiltonian~\cite{Rokhsar1988}, the coefficient of resonance terms is
$-t$. Here, it is changed to $it$, which could have qualitative
consequences~\cite{Ivanov2013}. Indeed, we shall demonstrate that the
topological order of the ground state of Eq. (\ref{eq:iRK}) is changed
to the double-semion order. Such RVB-type wave functions with
double-semion topological order could serve as candidate ground states
of frustrated spin-1/2 SU(2)-symmetric quantum magnets with frustrated
Heisenberg interactions. Naively, Eq.~\eqref{eq:iRK} seems to be not
invariant under the time reversal symmetry as $it$ goes to $-it$ under
complex conjugate $K$. However, since the quantum dimer model can be
viewed as the low-energy effective model in spin-singlet subspace,
time-reversal symmetry can be realized in a much more generic form
with $T=UK$, where $UU^*=1$ is a unitary transformation (the detailed
form of $U$ will be discussed later).  Actually, the time-reversal
symmetry becomes manifested in the low-energy effective-field theory
description for double-semion topological order, with
\begin{align}
\mathcal{L} = \frac{K_{IJ}^{\text{DS}}}{4\pi} \epsilon^{\lambda \mu \nu} a_{I\lambda} \partial_\mu a_{J\nu},\quad K^{\text{DS}}_{IJ} =  \left(
                 \begin{array}{cc}
                   2 & 0 \\
                   0 & -2 \\
                 \end{array}
               \right).
\end{align}

In Sec.~\ref{sec:sign-struct-dimer}, we construct the ground state
wave function of Eq.~\eqref{eq:iRK}. We show that under a proper local unitary transformation(which preserves the
topological order),
the ground state wave
function takes an extremely simple form and can be described by an equal-weight superposition of all dimer configurations
with signs of $\pm1$, and the signs of different dimer configurations
are determined by counting the number of loops in the corresponding
transition graph.

In Sec.~\ref{sec:monom-excit-stat}, we explicitly demonstrate that the
ground state we constructed has the double-semion topological
order. In particular, the monomer excitations obey the fusion rules and
statistics of the semion excitations. We first construct the sign
structure of the monomer wave function using a string operator which
we call a ``half-vison string.'' Using this result, the semionic
statistics of monomers is illustrated using the procedure first
proposed in Ref.~\onlinecite{Levin2003}, where the statistical phase of
exchanging two monomers is determined by comparing the Berry phases of
exchanging two monomers and moving one monomer along the same path.

Our construction of exactly solvable quantum dimer models with the
double-semion topological order can be generalized to other
two-dimensional nonbipartite lattices. In
Sec.~\ref{sec:gener-other-two}, we generalize our construction to the
star lattice (which is also known by many other names summarized in
Ref.~\onlinecite{Fjaerestad2008}) and to the kagome lattice. The star
lattice is an interesting playground to study the quantum dimer models
and spin-1/2 Kitaev models~\cite{Yao2007} as it has many nice
properties~\cite{Fjaerestad2008}, which are discussed in more detail
in Sec.~\ref{sec:gener-other-two}, and antiferromagnets on the star
lattice have been realized in a polymetic iron acetate
material~\cite{Zheng2007}. On the star lattice, both the model
Hamiltonian and its ground-state wave function are fully
symmetric. Furthermore, we demonstrate the double-semion topological
order of the ground-state wave function by mapping our model to the
string-net model~\cite{Levin2005} and to a symmetry-protected
topological (SPT) state in a spin model~\cite{Levin2012}. We also
generalize our construction to the kagome lattice.
Lastly, further discussions and the conclusion will be presented in
Sec.~\ref{sec:disc-concl}.

\section{Sign structures in dimer wave function}
\label{sec:sign-struct-dimer}
We start with a brief review of the exactly solvable RK Hamiltonian
and its ground-state wave function with the toric-code topological
order. On the triangular lattice, the RK Hamiltonian takes the following
form~\cite{sondhi2001}:
\begin{equation}
  \label{eq:HRK}
  \begin{split}
  H_{\text{RK}}=\sum_{\text{plaquette}}
      &-t\left(\ket{\plaq1}\bra{\plaq2}+\text{H.c.}\right)\nonumber\\
      &+v\left(\ket{\plaq1}\bra{\plaq1}
        +\ket{\plaq2}\bra{\plaq2}\right),
    \end{split}
\end{equation}
where the sum runs over all rhombic plaquettes. This Hamiltonian is
exactly solvable at $t=v$, known as the RK point, because there the
Hamiltonian can be rewritten as a sum of projection operators as
follows,
\begin{equation}
  \label{eq:HRKp}
    H_{\text{RK}}=t\sum_{\text{plaquette}}
      \left(\ket{\plaq1}-\ket{\plaq2}\right)
      \left(\bra{\plaq1}-\bra{\plaq2}\right).
\end{equation}
As a result, the wave function
\begin{equation}
  \label{eq:rk-wf}
  \ket{\Psi} = \sum_c\ket{c},
\end{equation}
which is an equal-amplitude superposition of all dimer configurations,
is an exact ground state of the Hamiltonian (here $c$ denotes a dimer
configuration). It can be shown that this wave function has the Z$_2$
topological order, which is the same as the toric-code model.

\begin{figure}[htbp]
  \centering
  \subfigure[\label{fig:sr:ref}]{\includegraphics[scale=0.7]{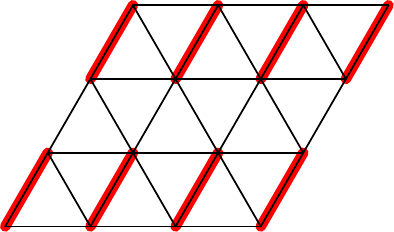}}
  \subfigure[\label{fig:sr:c}]{\includegraphics[scale=0.7]{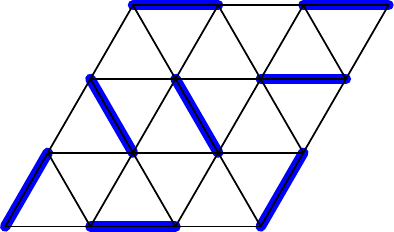}}
  \subfigure[\label{fig:sr:loop}]{\includegraphics[scale=0.7]{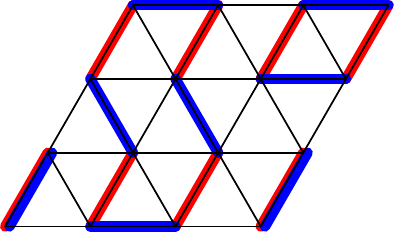}}
  \caption{Sign rule of dimer configurations. (a) The reference dimer
    configuration, which is a columnar VBS state. (b) An arbitrary
    dimer configuration. (c) The transition graph constructed by
    superposing the configurations in (a) and (b). In this transition
    graph, there are four loops, including two trivial length-two loops
    at the bottom-left and the bottom-right corners, and therefore the
    count $N_c=4$ in Eq.~\eqref{eq:sc}.}
  \label{fig:sr}
\end{figure}

In this work, we construct a dimer wave function with the double-semion
topological order. Motivated by the construction of wave functions
with such topological order in the string-net models~\cite{Levin2005},
we consider the following dimer wave function constructed as an
equal-weight superposition of different dimer configurations with a
sign structure:
\begin{equation}
  \label{eq:wf}
  |\Psi\rangle = \sum_cs(c)|c\rangle,
\end{equation}
where $s(c)=\pm1$ is a sign function defined as follows. First, we
construct a transition graph by superposing the dimer configuration
$c$ and a particular reference configuration $c_0$. Here we choose the
reference $c_0$ to be the columnar valence bond solid (VBS)
configuration shown in Fig.~\ref{fig:sr:ref}. Then the sign function
is determined from the number of transition loops $N_c$ in the
transition graph,
\begin{equation}
  \label{eq:sc}
  s(c) = (-1)^{N_c}.
\end{equation}
Note that in the transition graph, if the two dimer configurations
occupy the same bond, the two overlapping dimers form a trivial
length-two loop. This trivial loop also contributes a factor of $-1$ in
the wave function. The construction of transition loops is
demonstrated with an example in Fig.~\ref{fig:sr:loop}.

\begin{figure}[htbp]
  \centering
  \subfigure[\label{fig:sr:abc}]{\includegraphics{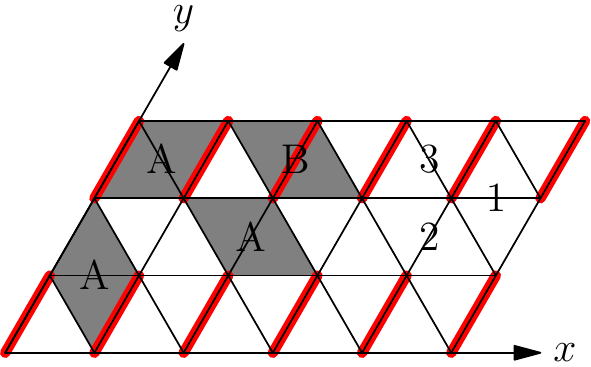}}\\
  \subfigure[\label{fig:tg:a}]{\includegraphics{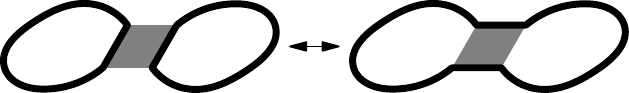}}\\
  \subfigure[\label{fig:tg:b}]{\includegraphics{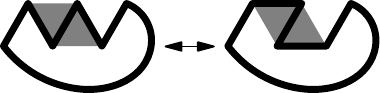}}\\
  \caption{Different types of resonance terms and their effects on
    transition graphs. The shaded rhombus shows the location of the
    resonance term, and the thick lines in (b) and (c) are transition
    loops. (a) Two types of resonance terms. The red dimers show the
    reference dimer configuration as in Fig.~\ref{fig:sr:ref}. (b)
    Effect of a type-A resonance term. A type-A term reconnects two
    transition loops and changes the number of loops by one. (c) Effect of a type-B
    resonance term. A type-B resonance term reconnects the loop
    locally and does not change the number of loops.}
  \label{fig:tg}
\end{figure}

Next, we show that the wave function in Eq.~\eqref{eq:wf} is the exact
ground state of the following extended RK Hamiltonian at the RK point
$t=v$:
\begin{widetext}
\begin{equation}
  \label{eq:eRK}
  \begin{split}
    H_{\text{RK}}^{\text{ext}}&=\sum\left[
      t\left(\ket{\plaq1}\bra{\plaq2}+\text{H.c.}\right)
      +v\left(\ket{\plaq1}\bra{\plaq1}
        +\ket{\plaq2}\bra{\plaq2}\right)\right]\\
    &+\sum\left[
     (-1)^y t\left(\ket{\plaq3}\bra{\plaq4}+\text{H.c.}\right)
      +v\left(\ket{\plaq3}\bra{\plaq3}
        +\ket{\plaq4}\bra{\plaq4}\right)\right]\\
    &+\sum\left[
      t\left(\ket{\plaqh5}\bra{\plaqh6}+\text{H.c.}\right)
      +v\left(\ket{\plaqh5}\bra{\plaqh5}
        +\ket{\plaqh6}\bra{\plaqh6}\right)\right],
  \end{split}
\end{equation}
\end{widetext}
where the three sums run over rhombic resonance plaquettes in three
different orientations respectively, and the sign of the resonance
terms with the second orientation alternates in even and odd rows, as
indicated by the factor of $(-1)^y$ [$y$ is the coordinate in the
$y$ direction, as labeled in Fig.~\ref{fig:sr:abc}]. This Hamiltonian
differs from the original RK Hamiltonian in Eq.~\eqref{eq:HRK} where
the signs of the resonance term on some plaquettes become
positive. For clarity, in the rest of the paper, we denote the
plaquettes with positive and negative resonance terms in
Eq.~\eqref{eq:eRK} as type-A and type-B plaquettes, respectively, as
shown in Fig.~\ref{fig:sr:abc}.

Similar to the original RK Hamiltonian, at the RK point $t=v$
this extended RK Hamiltonian can also be written as a sum of
projection operators on all plaquettes as follows,
\begin{equation}
  \label{eq:eRKp}
  \begin{split}
    H_{\text{RK}}^{\text{ext}}=&t\sum_A
      \left(\ket{\plaq1}+\ket{\plaq2}\right)
      \left(\bra{\plaq1}+\bra{\plaq2}\right)+\\
    &t\sum_B
      \left(\ket{\plaq3}-\ket{\plaq4}\right)
      \left(\bra{\plaq3}-\bra{\plaq4}\right),
 \end{split}
\end{equation}
where the two sums run over type-A and type-B plaquettes,
respectively. Hence, similar to the original RK Hamiltonian, the ground
state of this Hamiltonian is also a superposition of dimer
configurations that are annihilated by all projection operators in
Eq.~\eqref{eq:eRKp}. From the form of projection operators, we see that
the ground-state wave function should be an equal-amplitude
superposition of two different configurations resonating on a type-B
plaquette, but an equal-weight superposition with opposite signs of two
different configurations resonating on a type-A plaquette. We shall
argue that the sign function defined in Eq.~\eqref{eq:sc} by counting
the number of transition loops satisfies these requirements. To see
this, we first notice that a difference between type-A and type-B
plaquettes is that in a type-B plaquette, the dimer at the middle of
the plaquette belongs to the reference dimer configuration, while in a
type-A plaquette it does not, as shown in Fig.~\ref{fig:sr:abc}. As a
result, the two types of resonance terms act differently in the
transition graphs, as shown in Fig.~\ref{fig:tg}: A resonance term on
a type-A plaquette reconnects two transition loops into one (or splits
one into two), and therefore changes the parity of the total number of
transition loops. On the other hand, a resonance term on a type-B
plaquette alters a transition loop locally without changing the number
of transition loops. As a result, the wave function defined using the
sign in Eq.~\eqref{eq:sc} is indeed a ground state of the extended RK
Hamiltonian in Eq.~\eqref{eq:eRK} at $v=t$.

The global phase diagram of the extended RK Hamiltonian can be
obtained by mapping it to the original RK Hamiltonian via a (nonlocal)
unitary transformation that redefines the dimer basis using the sign
function,
\begin{equation}
  \label{eq:ut}
  \ket c\rightarrow \ket{\tilde c} = s(c)\ket c.
\end{equation}
After this (nonlocal) unitary transformation, the wave function in
Eq.~\eqref{eq:wf} becomes the original RK wave function in
Eq.~\eqref{eq:rk-wf} and, using the previous argument one can show
that the extended RK Hamiltonian in Eq.~\eqref{eq:eRK} becomes the
original RK Hamiltonian in Eq.~\eqref{eq:HRK}.  Thus, we show that the
wave function in Eq.~\eqref{eq:wf} is an exact ground state of the
extended RK Hamiltonian at $v=t$. Similar to the original RK
Hamiltonian, the extended Hamiltonian we study also has a first-order
phase transition at $v=t$, separating a staggered valence bond solid
phase at $v>t$ and a gapped liquid phase at $v<t$, which is smoothly
connected to the exact wave function at $v=t$ in
Eq.~\eqref{eq:wf}. Hence, in this work, we use this wave function as a
representative wave function to study the gapped liquid phase and the
topological order therein. We also stress that the
$H_{\text{RK}}^{\text{ext}}$ and $H_{\text{RK}}$ indeed have different
topological orders since Eq.~\eqref{eq:ut} is not a \emph{local}
unitary transformation.

Finally, we show that $H_{\text{RK}}^{\text{ext}}$ and $H$ have the
same topological order since they can be mapped to each other through
a \emph{local} unitary transformation, defined by
$b_{ij}\rightarrow b_{ij}e^{i\theta_{ij}}$, where the bosonic operator
$b_{ij}$ (with $b_{ij}^2=0$) annihilates the dimer excitation on the
bond $(ij)$, and $\theta_{ij}$ is a phase factor that depends on the
bond.  As shown in Fig.~\ref{fig:sr:abc},
here we assign $\theta_{ij}=0$ for bonds with the first
orientation, $\theta_{ij}=\pi/4$ for bonds with the second
orientation, and $\theta_{ij}=\pi/4 ~(3\pi/4)$ for bonds with the
third orientation on even (odd) rows, respectively.
With this local unitary
transformation, one can check that the Hamiltonian defined in
Eq.~\eqref{eq:eRK} becomes the one we have introduced at the beginning of
the paper in Eq.~\eqref{eq:iRK}, and the wave function defined in
Eq.~\eqref{eq:wf} becomes
\begin{equation}
  \label{eq:wf2}
  |\Psi_0\rangle = \sum_c(-1)^{N_c}
  e^{i\frac\pi4 N_2}
  e^{i\frac\pi4 N_3^{\text{even}}}
  e^{i\frac{3\pi}4 N_3^{\text{odd}}}|c\rangle,
\end{equation}
where $N_2$ denotes the total number of dimers in the second
orientation, and $N_3^{\text{even}}$ ($N_3^{\text{odd}}$) denotes the
number of dimers in the third orientation on even (odd) rows,
respectively. This local unitary transformation restores the
translation and rotation symmetries in $H$. In addition, the
ground-state wave function of $H$ also preserves translation and
rotation symmetries because, using the (nonlocal) unitary
transformation in Eq.~\eqref{eq:ut}, we can show that $|\Psi_0\rangle$
is the unique ground state (within a topological section) of the
Hamiltonian, which itself is invariant under these symmetry
transformations.

Since $H_{\text{RK}}^{\text{ext}}$ is real and manifestly
time-reversal invariant, the above local unitary transformation
implies the time-reversal symmetry operator of $H$ should be defined
by $T=UK$ with $U:b_{ij}\rightarrow b_{ij}e^{i\theta_{ij}}$, which is
different from the physical time-reversal operation if the dimers are
interpreted as spin-singlet pairs in a spin model. Therefore, in this
context, the model in Eq.~\eqref{eq:iRK} breaks time-reversal symmetry
(and also inversion symmetry). However, such an antiunitary operation
may be realized as the physical time-reversal operation if the dimers
are interpreted in other physical contexts.

On the other hand, we note that the Hamiltonian in Eq.~\eqref{eq:eRK}
explicitly breaks symmetries of the triangular lattice because the
local unitary transformation
$e^{i\frac\pi4 N_2} e^{i\frac\pi4 N_3^{\text{even}}}e^{i\frac{3\pi}4
  N_3^{\text{odd}}}$
does not preserves these symmetries. In particular, it breaks the
six-fold rotational symmetry and the translational symmetry along the
$y$ direction, as the resonance terms have different signs in
different orientations and on different rows. The wave function
defined in Eq.~\eqref{eq:wf} also lacks these lattice symmetries, as
its definition is based on a specific reference dimer configuration
that breaks them. However, this is not essential to our study of the
topological order in the state, as the intrinsic topological order we
are interested in requires no symmetry to protect it.

In summary, in this section, we construct a wave function using a sign
structure determined by the number of transition loops in the
transition graph against a particular reference dimer
configuration. This wave function is constructed as the ground state
of an exactly solvable RK-like Hamiltonian, given by
Eq.~\eqref{eq:iRK}.  We construct two forms for the exactly solvable
Hamiltonian and its ground-state wave function, which are related by a
local unitary transformation. The first form in Eqs.~\eqref{eq:eRK} and
\eqref{eq:wf} breaks rotation and translation symmetries, while
the second form in Eqs.~\eqref{eq:iRK} and \eqref{eq:wf2} preserves
all of the lattice symmetries but with twisted time-reversal and mirror
symmetries. Despite the different symmetries, the two ground-state
wave functions have the same intrinsic topological order, which does
not need the protection of any global symmetry. Although $H$ appears
to be simpler, $H_{\text{RK}}^{\text{ext}}$ has a simpler ground-state
wave function. Hence, in Sec.~\ref{sec:monom-excit-stat} we use
$H_{\text{RK}}^{\text{ext}}$ to discuss the intrinsic topological
order in the ground state.

\section{Monomer excitations and their statistics}
\label{sec:monom-excit-stat}

Now we study the topological order in the ground-state
wave function introduced in the previous section by investigating
the topological excitations above the ground state. In particular, we
consider the monomer excitations, which are sites that have no dimer
connected to them. In the dimer liquid state with the toric-code
topological order, the monomer excitations are themselves bosons but
have a nontrivial mutual statistics with another type of topological
excitations called the visons. Here we shall show that for the dimer
liquid state described by the wave function in Eq.~\eqref{eq:wf}, the
monomer excitations obey nontrivial semionic statistics. The
statistics of the topological excitations implies that the ground
state features the so-called double-semion topological order instead
of the toric-code order in the original RK wave function.

\begin{table}[tbp]
  \centering
  \caption{Self- and mutual statistics of topological excitations in
    the double-semion topological order. $s$, $\bar s$, and $b$ denotes
    the semion, the antisemion and the bosonic bound state, respectively.}
  \label{tab:stat}
  \begin{tabular*}{\columnwidth}{@{\extracolsep{\fill}}cccc}
    \hline\hline
    $e^{i\theta}$ & $s$ & $\bar s$ & $b$ \\
    \hline
    $s$ & $i$ & 0 & $-1$ \\
    $\bar s$ & 0 & $-i$ & $-1$ \\
    $b$ & $-1$ & $-1$ & 0\\
    \hline\hline
  \end{tabular*}
\end{table}

\begin{table}[tbp]
  \centering
  \caption{Fusion rules between topological excitations in the
    double-semion topological order.}
  \label{tab:fusion}
  \begin{tabular*}{\columnwidth}{@{\extracolsep{\fill} }cccc}
    \hline\hline
    $\times$ & $s$ & $\bar s$ & $b$ \\
    \hline
    $s$ & 1 & $b$ & $\bar s$ \\
    $\bar s$ & $b$ & 1 & $s$ \\
    $b$ & $\bar s$ & $s$ & 1\\
    \hline\hline
  \end{tabular*}
\end{table}

We begin with a brief review of the topological excitations in the
double-semion topological order. In such a state, there are three
types of excitations: the semion, the antisemion and the bosonic bound
state of two semions. The statistics and fusion rules of these
excitations are summarized in Tables~\ref{tab:stat} and
\ref{tab:fusion}, respectively. We shall see that in the extended
quantum dimer models, these three types of excitations correspond to
two types of monomers and the vison, respectively.

\begin{figure}[htbp]
  \centering
  \subfigure[\label{fig:string:a}]{\includegraphics[scale=0.53]{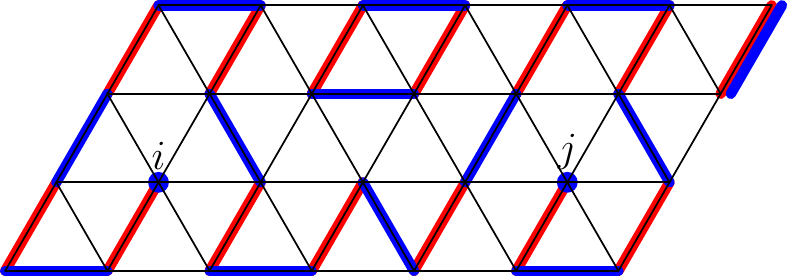}}
  \subfigure[\label{fig:string:b}]{\includegraphics[scale=0.53]{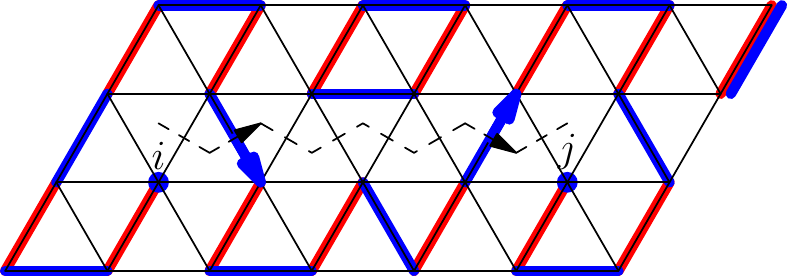}}
  \caption{A monomer string and a half-vison string. (a) The
    transition graph of a dimer configuration with two monomer
    excitations. Besides the transition loops, the graph also contains
    a string connecting two monomer sites, for which we coin the name
    ``monomer string.'' (b) A half-vison string connecting the two
    monomer sites. The half-vison string is represented by a dashed
    line. Both the monomer string and the half-vison string are
    oriented from site $i$ to $j$. The arrows on the dimers and the
    dashed line show the direction of monomer and half-vison
    strings. Note that the half-vison string runs through the dual
    lattice, so it starts and ends on an adjacent dual lattice site to
    the monomer excitations. According to the definition of the
    half-vison string, the two crossings at the left and the right
    gives phase factors of $-i$ and $i$, respectively. This is
    explained in Appendix~\ref{sec:semion-string}.}
  \label{fig:string}
\end{figure}

First, we consider the dimer wave function that contains monomer
excitations at fixed locations. The monomer wave function is an
eigenstate of the Hamiltonian in Eq.~\eqref{eq:eRK} in the Hilbert
space that consists of dimer configurations that have no dimer connecting
to the monomer sites. We start with the simple case that has two
monomers at sites $i$ and $j$, and consider a wave function that is a
superposition of different dimer configurations,
\begin{equation}
  \label{eq:wf-ij}
  \ket{\Psi(i, j)}=\sum_{c[i,j]}s(c[i,j])\ket{c[i,j]},
\end{equation}
where $c[i,j]$ denotes dimer configurations with two monomer defects
at sites $i$ and $j$. To determine the phase factor $s(c[i,j])$, we
again consider the transition graph between the configuration $c[i,
j]$ and the reference state in Fig.~\ref{fig:sr:ref}. Since the
configuration $c[i,j]$ has two monomer defects, besides the transition
loops the transition graph also contains a string connecting the two
defect sites $i$ and $j$, as illustrated with an example in
Fig.~\ref{fig:string:a}. In this paper, we coin the name ``monomer
string'' for this string. Hence naturally in the phase factor $s(c[i,
j])$, besides the factor of $-1$ contributed from each transition loop,
there is an extra phase factor depending on the configuration of the
monomer string,
\begin{equation}
  \label{eq:scij-l}
  s(c[i, j]) = (-1)^{N_c} \phi(M_{ij}),
\end{equation}
where $M_{ij}$ denotes the monomer string connecting $i$ and $j$.

\begin{figure}[htbp]
  \centering
  \subfigure[\label{fig:stg:b}]{\includegraphics{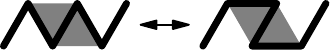}}~~
  \subfigure[\label{fig:stg:at}]{\includegraphics{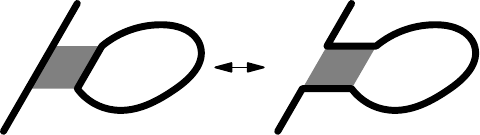}}\\
  \subfigure[\label{fig:stg:a}]{\includegraphics{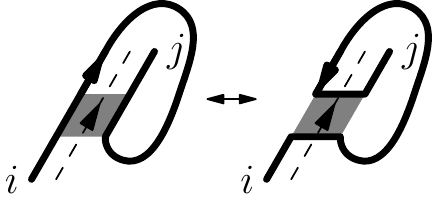}}
  \caption{Effects of different types of resonance terms on the defect
    string connecting two monomers. Similar to Fig.~\ref{fig:tg}, the
    shaded area shows the location of the resonance term and the thick
    line shows the defect string that connects two monomer
    excitations. The dashed line shows the half-vison string.  (a) A
    type-B resonance term changes the shape of the string locally
    without changing the topology of the string. (b) A type-A
    resonance term absorbs a nearby loop into the string. (c) A type-A
    resonance term reconnects the monomer string. In the left
    configuration the half-vison string does not intersect the monomer
    string, and in the right configuration the half-vison string
    intersects the monomer string twice with the same phase
    factors $i$, so the total phase is $-1$. Therefore, the half-vison
    string gives an extra phase factor $-1$ after the resonance.}
  \label{fig:stg}
\end{figure}

To determine the form of $\phi$, we study how the resonance terms in
the extended RK Hamiltonian act on the monomer string. The type-B
terms act trivially on the monomer string by changing the shape of
the string locally, as shown in Fig.~\ref{fig:stg:b}, while it does not change the sign of the wave function. The
type-A terms, on the other hand, act in two ways on the monomer
string. First, in Fig.~\ref{fig:stg:at}, the action of such resonance
term merges the string with a nearby loop. In this process the wave
function changes sign because the type-A term is positive in the
Hamiltonian, while the count of loop number $N_c$ also changes by
one. Therefore, the factor $\phi(M_{ij})$ is unchanged in this
resonance process. Second, in Fig.~\ref{fig:stg:a}, it is illustrated
that after the action of a type-A resonance term, the string is
reconnected with itself and no transition loop is created or
annihilated. Since the type-A term is positive in the Hamiltonian, and
$N_c$ stays the same, the phase factor $\phi(M_{ij})$ must acquire a
minus sign. Such a phase factor can be constructed using another
string operator connecting the two monomer sites through the dual
lattice. This string operator acts like a square root of the vison
string operator; therefore, we coin the name ``half-vison string.''

The definition of a half-vison string operator is illustrated in
Fig.~\ref{fig:string:b}. Here we only give a simplified version of its
definition for the special case that there are only two monomers and
the half-vison string connects the two monomer sites. (Since the
half-vison string goes through the dual lattice, when we say the
half-vison string connects a monomer site we mean the string
terminates at a nearest-neighbor dual lattice site around the monomer
site. In Appendix~\ref{sec:semion-string}, we discuss this in more
detail and show that this arbitrary choice does not affect our
result.) The definition for general cases is given in
Appendix~\ref{sec:semion-string}. We first assign a direction to the
monomer string in the transition graph, and the half-vison string is
also oriented. The action of the string operator is then defined as a
product of phase factors everytime the half-vison string and the
monomer string intersect. The phase factor at the intersection point
depends on the angle between the dimer and the half-vison string: the
phase factor is $i$ if the dimer runs from the right to the left
relative to the direction of the half-vison string, and $-i$
otherwise. Some basic properties of the half-vison string operator are
discussed in Appendix~\ref{sec:semion-string}: Similar to the vison
string operator, it is also independent of the choice of the path and
effectively defines two local operators at the two ends of the
string. Moreover, two parallel half-vison strings fuse into one vison
string, which is the reason we chose its name.

\begin{figure}[htbp]
  \centering
  \subfigure[\label{fig:fusion:a}]{\includegraphics[scale=0.8]{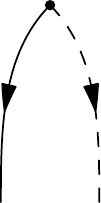}}~~~~~
  \subfigure[\label{fig:fusion:b}]{\includegraphics[scale=0.8]{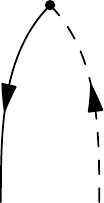}}~~~~~
  \subfigure[\label{fig:fusion:c}]{\includegraphics[scale=0.8]{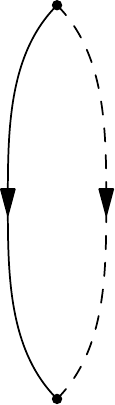}}~~~~~
  \subfigure[\label{fig:fusion:d}]{\includegraphics[scale=0.8]{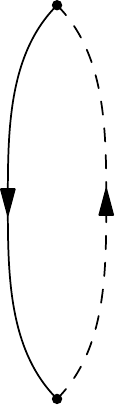}}~~~~~
  \subfigure[\label{fig:fusion:e}]{\includegraphics[scale=0.8]{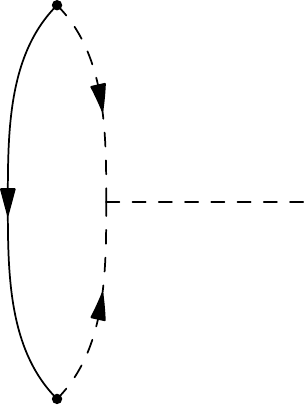}}
  \caption{Monomer excitations and fusion rules. The solid lines with
    arrows are the monomer strings, and dashed lines with arrows are
    the half-vison strings. The dashed line without an arrow is a
    vison string. (a) A monomer excitation with one outgoing monomer
    string and one outgoing half-vison string. (b) An antimonomer with
    one outgoing monomer string and one incoming half-vison
    string. (c) Two monomers are joined by one monomer string and one
    half-vison string running in parallel directions and fuse into
    vacuum. (d) Two antimonomers are joined by one monomer string and
    one half-vison string running in opposite directions and fuse into
    vacuum. (e) One monomer and one antimonomer are joined by one
    monomer string, but the two half-vison strings meet in the middle
    and fuse into a vison string. Hence one monomer and one
    anti-monomer fuse into a vison.}
  \label{fig:fusion}
\end{figure}

The phase factor $\phi(M_{ij})$ in the two-monomer wave function can
be chosen to be the phase factor of a half-vison string operator that
connects the two monomers,
\begin{equation}
  \label{eq:phi}
  \phi(M_{ij}) = H_{i\rightarrow j}(M_{i\rightarrow j}),
\end{equation}
where $H_{i\rightarrow j}$ denotes a half-vison string operator from
$i$ to $j$, and $M_{i\rightarrow j}$ denotes the oriented monomer
string. To prove that the wave function constructed with this phase
factor is indeed an eigenstate of the Hamiltonian in
Eq.~\eqref{eq:eRK}, we need to show that the phase factor in
Eq.~\eqref{eq:phi} is invariant under the resonance processes shown in
Figs.~\ref{fig:stg:b} and \ref{fig:stg:at}, but changes sign in
Fig.~\ref{fig:stg:a}. Since the half-vison string operator is
path independent, we can choose its path to simplify our
discussion. For the resonance processes in Figs.~\ref{fig:stg:b} and
\ref{fig:stg:at}, we choose the path of the half-vison string
such that it does not intersect with the piece of monomer string shown
in the figures. Therefore, the phase factor in Eq.~\eqref{eq:phi} is
indeed unchanged. For the resonance process in Fig.~\ref{fig:stg:a},
we let the half-vison string cut through the resonance plaquette, as
shown in that figure. It is easy to check that after the resonance, the
phase factor $\phi(M_{ij})$ acquires a minus sign, which is consistent
with the Hamiltonian. Therefore, the two-monomer wave function defined
using $\phi(M_{ij})$ in Eq.~\eqref{eq:phi} is indeed an eigenstate of
the Hamiltonian in Eq.~\eqref{eq:eRK}.

Similarly, one can construct another wave function using a half-vison
string running in the opposite direction,
\begin{equation}
  \label{eq:phi2}
  \phi^\ast(M_{ij}) = H_{j\rightarrow i}(M_{i\rightarrow j}),
\end{equation}
and using the definition of half-vison string it is easy to see that
this phase factor is the complex conjugate of the one in
Eq.~\eqref{eq:phi}. In this construction, the monomer excitations can
be viewed as a bound state of one end of a monomer string and one end
of a half-vison string. Because both types of strings are oriented,
the monomer belongs to different types depending on the direction of
the strings. Since the wave function is invariant if both the monomer
string and the half-vison string are reversed, there are two types of
monomer excitations, with the two strings in the same or opposite
directions, respectively. We denote the bound state of two starting or
two ending points of the two strings as the monomer, and the bound
state of one starting and one ending point of the two strings as the
antimonomer, respectively, as illustrated in Figs.~\ref{fig:fusion:a}
and \ref{fig:fusion:b}. Using this convention, the wave function
$\ket{\Psi(i, j)}$ contains two monomer excitations, whereas its
complex conjugate $\ket{\Psi^\ast(i, j)}$ contains two antimonomers.

\begin{figure}[htbp]
  \centering
  \subfigure[\label{fig:hex:a}]{\includegraphics{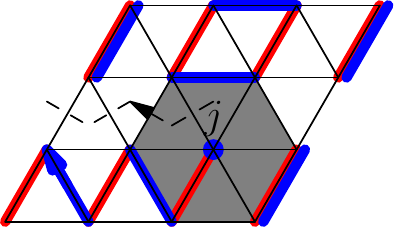}}
  \subfigure[\label{fig:hex:b}]{\includegraphics{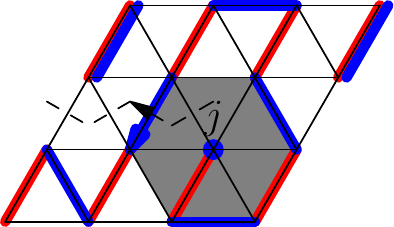}}
  \caption{Length-six resonance term.}
  \label{fig:hex}
\end{figure}

In the two-monomer Hilbert space, there are two degenerate ground-state
wave functions $\ket{\Psi(i, j)}$ and $\ket{\Psi^\ast(i, j)}$,
representing states with two monomers or two antimonomers,
respectively. This double degeneracy of two-monomer states is
accidental and not protected by the topological order. In particular, it
can be lifted by local perturbations near the monomer
excitations~\cite{Lan2013}. As an example, we consider a resonance term
on a length-six loop surrounding one monomer excitation, as shown in
Fig.~\ref{fig:hex}. For the configuration shown in this example, the
action of the resonance term annihilates one transition loop and
creates one intersection between the half-vison string and the monomer
string with the factor $i$. Hence the sign of the two-monomer wave
function $\ket{\Psi(i, j)}$ changes by $-i$. On the other hand, the
two-antimonomer wave function $\ket{\Psi^\ast(i, j)}$ changes by
$i$. Therefore, adding such a resonance term with coefficient $\mp i$
to the Hamiltonian lifts the degeneracy and favors the state with two
monomers and two antimonomers, respectively, as this term breaks the
time-reversal symmetry which relates the two wave functions
$\ket{\Psi(i, j)}$ and $\ket{\Psi^\ast(i, j)}$.

Using the construction of the wave function, we can check the fusion
rules between monomer, antimonomer, and vison excitations, as
illustrated in Fig.~\ref{fig:fusion}. First, if we have two monomers,
they can be joined by one monomer string and one half-vison string
running in the same direction, and therefore they fuse into a trivial
state. Likewise, two antimonomers can be joined by one monomer string
and one half-vison string running in opposite directions, and also fuse
into a trivial state. On the other hand, if we have one monomer and
one antimonomer, they can be joined by one monomer string, but there
are two half-vison strings coming out from them. The two half-vison
strings meet and they can fuse into one vison string (see
Appendix~\ref{sec:semion-string}). Therefore, one monomer and one
antimonomer fuse into a vison.

\begin{figure}[htbp]
  \centering
  \subfigure[\label{fig:exchange:a}]{\includegraphics[scale=0.7]{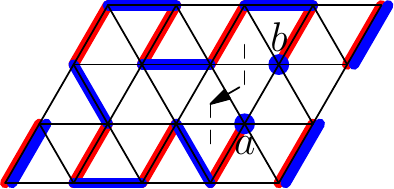}}
  \subfigure[\label{fig:exchange:b}]{\includegraphics[scale=0.7]{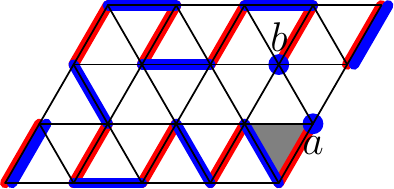}}
  \subfigure[\label{fig:exchange:c}]{\includegraphics[scale=0.7]{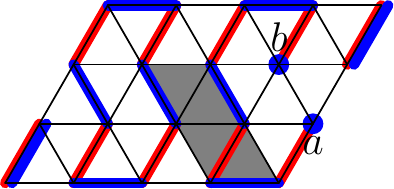}}
  \subfigure[\label{fig:exchange:d}]{\includegraphics[scale=0.7]{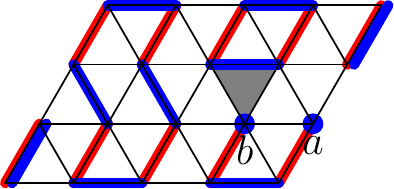}}
  \subfigure[\label{fig:exchange:e}]{\includegraphics[scale=0.7]{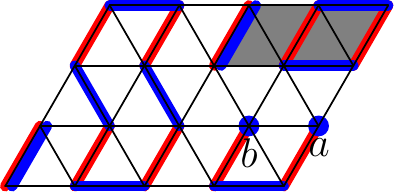}}
  \subfigure[\label{fig:exchange:f}]{\includegraphics[scale=0.7]{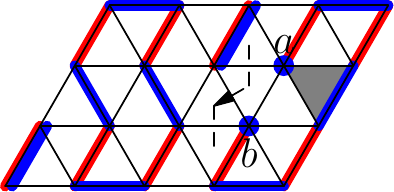}}
  \caption{Exchanging two monomers.}
  \label{fig:exchange}
\end{figure}

Next, we explicitly demonstrate the nontrivial semionic statistics of
monomer excitations by calculating the exchange algebra between two
monomer excitations. This can be computed by first calculating the
Berry phase acquired by exchanging two monomers, then subtracting the
Berry phase acquired by moving one monomer along the same
path~\cite{Levin2003}.

\begin{figure}[htbp]
  \centering
  \subfigure[\label{fig:hopref:a}]{\includegraphics[scale=0.85]{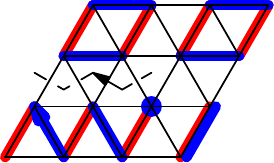}}
  \subfigure[\label{fig:hopref:b}]{\includegraphics[scale=0.85]{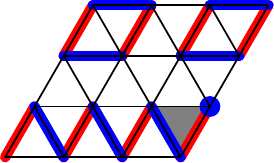}}
  \subfigure[\label{fig:hopref:c}]{\includegraphics[scale=0.85]{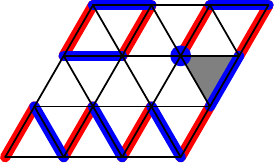}}
  \subfigure[\label{fig:hopref:d}]{\includegraphics[scale=0.85]{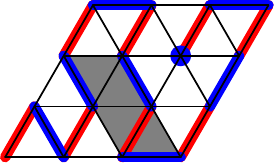}}
  \subfigure[\label{fig:hopref:e}]{\includegraphics[scale=1.0]{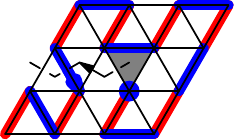}}
  \caption{Moving one monomer along exactly the same path as in
    Fig.~\ref{fig:exchange}.}
  \label{fig:hopref}
\end{figure}

As a demonstration, we consider the exchange process on the dimer
configurations shown in Fig.~\ref{fig:exchange} and corresponding
hopping process shown in Fig.~\ref{fig:hopref}. The Berry phase
accumulated through the exchange process is determined by multiplying
the signs of the hopping and resonance terms used in the process, and
the relative phase between the initial and final dimer
configurations. In the exchange process, we used three hopping terms,
which exchange the location of the monomer with a nearby dimer in the
shaded triangle. Here the signs of the hopping terms can be neglected,
because the same hopping terms are also used in Fig.~\ref{fig:hopref}
and the signs will cancel when subtracting the two Berry phases. From
the Hamiltonian in Eq.~\eqref{eq:eRK}, we see that the product of the
signs of the resonance terms used in Fig.~\ref{fig:exchange} is
$+1$. The relative phase between the initial and final dimer
configurations in Figs.~\ref{fig:exchange:a} and \ref{fig:exchange:b}
is $-1$, since one transition loop is created during the process, and
the half-vison string contributes no factor in both
Figs.~\ref{fig:exchange:a} and \ref{fig:exchange:f}. Therefore, the
total Berry phase acquired in Fig.~\ref{fig:exchange} is $-1$. The
Berry phase acquired in the hopping process in Fig.~\ref{fig:hopref}
can be determined similarly. Here the total sign of the resonance
terms is again $+1$, and according to the sign of the monomer wave
function defined in Eq.~\eqref{eq:wf-ij}, the relative sign between the
final state in Fig.~\ref{fig:hopref:e} and the initial state in
Fig.~\ref{fig:hopref:a} is $i$, as the parity of the number of loops
is the same, but the half-vison string cuts the monomer string once
with a phase factor $i$ in Fig.~\ref{fig:hopref:e}. Therefore, the
total Berry phase is $i$. Subtracting this from the Berry phase in
Fig.~\ref{fig:exchange}, we conclude that the statistical phase
accumulated by exchanging two monomers is $i$, which is consistent
with exchanging two semions as listed in Table~\ref{tab:stat}.

Similarly, if we were exchanging two antimonomers instead, the
statistical phase determined by the previous argument would become
$-i$ because the winding angle in Fig.~\ref{fig:hopref:a} becomes
$-2\pi$ as the monomer string reverses its direction. Therefore, we
conclude that the monomer and antimonomer excitations obey the
statistics and fusion rules of semion and antisemion excitations in
the double-semion topological order, respectively. This implies that
the gapped phase represented by the wave function in Eq.~\eqref{eq:wf}
indeed has the double-semion topological order.

\section{Generalization to the star lattice}
\label{sec:gener-other-two}

Our construction of the dimer wave function with the double-semion
topological order and the corresponding exactly solvable Hamiltonian
can be easily generalized to other two-dimensional nonbipartite
lattices, such as the star lattice and the kagome lattice. In particular,
when generalized to the star lattice, the constructed wave function
and the corresponding exactly solvable quantum dimer Hamiltonian are
invariant under both time-reversal and lattice symmetries in an
obvious way. We also briefly discuss the generalization to the kagome
lattice.

\begin{figure}[htbp]
  \centering
  \includegraphics{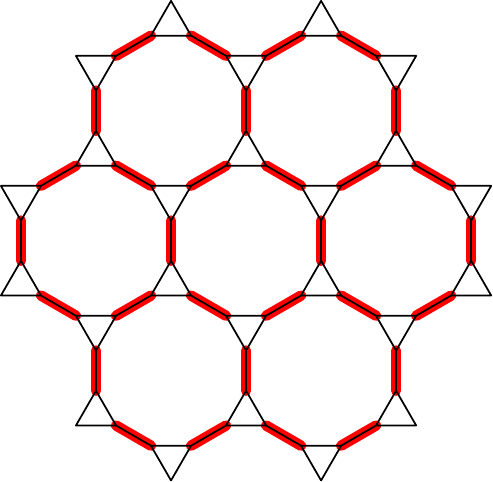}
  \caption{Fully symmetric dimer configuration on the star lattice.}
  \label{fig:star}
\end{figure}

The dimer wave function constructed in Eq.~\eqref{eq:wf} for the
triangular lattice breaks lattice symmetries because the choice of the
reference dimer configuration breaks them. However, on the star
lattice, a fully symmetric dimer configuration exists, as shown in
Fig.~\ref{fig:star}. Using this configuration as the reference, the
wave function constructed in Eq.~\eqref{eq:wf} becomes fully
symmetric.

Furthermore, an exactly solvable Hamiltonian can be constructed such
that the wave function in Eq.~\eqref{eq:wf} is its ground
state. Similar to the kagome lattice, on the star lattice the dimer
configurations also have arrow and pseudospin
representations~\cite{Elser1993,Misguich2002,Fjaerestad2008}, and from
these representations an exactly solvable Hamiltonian whose ground
state is the RK wave function is constructed as
follows~\cite{Misguich2002,Fjaerestad2008}:
\begin{equation}
  \label{eq:HRKstar}
  H=-\sum_D\sum_\alpha\left[
    |L_\alpha(D)\rangle\langle\bar L_\alpha(D)+\text{H.c.}\right],
\end{equation}
where the sum runs over all dodecagons labeled by $D$, and on each
dodecagon the sum runs over all even-length resonance loops
surrounding it. For each dodecagon, there are 32 such loops, denoted by
$\alpha=1,\ldots,32$. For each loop $\alpha$ on the dodecagon $D$,
$L_\alpha$ and $\bar L_\alpha$ denote the two different dimer
configurations that form the resonance loop. It is shown in
Ref.~\onlinecite{Fjaerestad2008} that the ground state of this
Hamiltonian is the RK wave function in Eq.~\eqref{eq:rk-wf} which supports Z$_2$ topological order. To obtain
the Hamiltonian for the double-semion wave function in
Eq.~\eqref{eq:wf}, we use the unitary transformation in
Eq.~\eqref{eq:ut}. After this unitary transformation, the RK wave
function in Eq.~\eqref{eq:rk-wf} is mapped to the double-semion wave
function in Eq.~\eqref{eq:wf}, and the Hamiltonian in
Eq.~\eqref{eq:HRKstar} is mapped to the following form:
\begin{equation}
  \label{eq:HRKstar-ds}
  H=-\sum_D\sum_\alpha \left[(-1)^{\frac{l(\alpha)}2+1}
    |L_\alpha(D)\rangle\langle\bar L_\alpha(D)|+\text{H.c.}\right],
\end{equation}
where $l(\alpha)$ denotes the length of the resonance loop (which is always
even).

The double-semion topological order in the ground state of the
Hamiltonian in Eq.~\eqref{eq:HRKstar-ds} can be understood by
repeating the argument in Sec.~\ref{sec:monom-excit-stat} on the
star lattice. Moreover, it can also be shown using dualities to two
previously studied models with the double-semion topological
order. First, this dimer model on the star lattice can be mapped to
a loop model, which is a special case of the general string-net models
discussed in Ref.~\onlinecite{Levin2005}. Because the star lattice is
trivalent (each lattice site has three nearest neighbors), a dimer
configuration on such a lattice can be mapped to a loop
configuration~\cite{Yao2010}, which occupies the bonds that are not
occupied by the dimers. Under this mapping, the original RK wave
function is mapped to an equal-amplitude superposition of all loop
configurations, which carries the toric-code topological
order~\cite{Levin2005}. On the other hand, the wave function in
Eq.~\eqref{eq:wf} is mapped to an equal-weight superposition of loop
configurations with a sign structure. Using the Hamiltonian in
Eq.~\eqref{eq:HRKstar-ds}, we can show that the ground-state wave
function in the loop model has the following form:
\begin{equation}
  \label{eq:wfl}
  |\Psi\rangle = \sum_X(-1)^{X_c}|X\rangle,
\end{equation}
where $X$ denotes all possible loop configurations, and $X_c$ is the
count of loops in a configuration. As shown in
Ref.~\onlinecite{Levin2005}, this wave function has the double-semion
topological order. This implies that the corresponding dimer wave
function also has this nontrivial topological order.

Second, the quantum dimer model in Eq.~\eqref{eq:HRKstar-ds} can be
mapped to a symmetry-protected topological (SPT) state discussed in
Ref.~\onlinecite{Levin2012}. Using the pseudospin representation,
quantum dimer models on the star lattices can be mapped to a spin
model~\cite{Fjaerestad2008} with pseudospin degrees of freedom that
locate at the centers of the dodecagon and form a triangular
lattice. In this mapping, the domain walls in the spin configuration
correspond to the loops in the transition graph against an arbitrary
reference dimer configuration (in our case, the reference configuration
can be chosen as the state in Fig.~\ref{fig:star}). The model in
Eq.~\eqref{eq:HRKstar} is mapped to the following spin model:
\begin{equation}
  \label{eq:hps}
  H=-\Gamma\sum_D\sigma^x(D),
\end{equation}
where $\sigma^x(D)$ is the Pauli matrix operator that acts on the
pseudospin. The ground state of this pseudospin Hamiltonian is a
trivial paramagnetic state. If we map this pseudospin model back to a
quantum dimer model, the mapping between the pseudospin and dimer
states is two to one, where two spin configurations related by the
operation $\prod_D\sigma^x(D)$ are mapped to the same dimer
configuration. Therefore, the operation $\prod_D\sigma^x(D)$ is a Z$_2$
global symmetry in the pseudospin model and becomes a gauge symmetry
in the quantum dimer model. In Ref.~\onlinecite{Levin2012}, it is shown
that gauging a Z$_2$ symmetry in a trivial paramagnetic state gives a
state with the toric-code topological order, which is consistent with
the intrinsic topological order carried by the original RK wave
function. Next, we consider mapping the quantum dimer model in
Eq.~\eqref{eq:HRKstar-ds} to a spin model in the pseudospin
representation. Since the loops in the transition graph are mapped to
domain walls in the pseudospin configuration, the total number of
loops $N_c$ in the sign structure of the wave function in
Eq.~\eqref{eq:wf} is mapped to the total number of domain
walls. Therefore, the ground-state wave function in Eq.~\eqref{eq:wf}
is mapped to the spin wave function discussed in
Ref.~\onlinecite{Levin2012},
\begin{equation}
  \label{eq:wfspt}
  |\Psi\rangle = \sum_{\{\sigma_D^z\}}(-1)^{N_{\text{dw}}}
  |\{\sigma_D^z\}\rangle,
\end{equation}
where $|\{\sigma_D^z\}\rangle$ denotes an Ising basis, and
$N_{\text{dw}}$ counts the number of domain walls in the spin
configuration. Correspondingly, the Hamiltonian in
Eq.~\eqref{eq:HRKstar-ds} is mapped to the following
form~\cite{Levin2012}:
\begin{equation}
  \label{eq:hpsspt}
  H=-\Gamma\sum_DB_D,\quad
  B_D=-\sigma_D^x\prod_{\langle DD_1D_2\rangle}
  i^{\frac{1-\sigma_{D_1}^z\sigma_{D_2}^z}2},
\end{equation}
where the product in the definition of $B_D$ runs over the six
triangles $\langle DD_1D_2\rangle$ containing the site $D$. In
Ref.~\onlinecite{Levin2012}, it is shown that the ground state in
Eq.~\eqref{eq:wfspt} has a nontrivial SPT order, protected by the
aforementioned global Z$_2$ symmetry. Furthermore, gauging this Z$_2$
symmetry gives a state with the double-semion topological order. This
is consistent with our conclusion that the dimer state in
Eq.~\eqref{eq:wf} carries the double-semion topological order.

Similarly, this construction can be implemented on the kagome lattice,
where the Z$_2$ RK wave function is the ground state of an exactly
solvable Hamiltonian~\cite{Misguich2002} analogous to
Eq.~\eqref{eq:HRKstar},
\begin{equation}
  \label{eq:HRKkagome}
  H=-\sum_H\sum_\alpha\left[
    |L_\alpha(H)\rangle\langle\bar L_\alpha(H)+\text{h. c.}\right],
\end{equation}
where $H$ iterates over all hexagons in the kagome lattice, and
$\alpha$ labels the even-length resonance loops surrounding a
hexagon. On a kagome lattice, the double-semion dimer wave function in
Eq.~\eqref{eq:wf} can be constructed using a particular reference
dimer configuration. Unlike the star lattice, such reference
configuration always breaks translational symmetries of the kagome
lattice, and as a result the constructed wave function and the
corresponding exactly solvable Hamiltonian also lacks lattice
symmetries. Again using the unitary transformation in
Eq.~\eqref{eq:ut} we can obtain the general form of the exactly
solvable Hamiltonian whose ground state is the double-semion dimer
wave function,
\begin{equation}
  \label{eq:HRKkagome-ds}
  H=-\sum_{H,\alpha} \left[(-1)^{\frac{l(\alpha)}2+r(H,\alpha)+1}
    |L_\alpha(H)\rangle\langle\bar L_\alpha(H)|+\text{H.c.}\right],
\end{equation}
where $r(H,\alpha)$ counts the number of reference dimers which
connect next-nearest-neighbor sites in the resonance loop $\alpha$
around the hexagon $H$. We note that this Hamiltonian explicitly
breaks lattice symmetries because of the factor $r(H,\alpha)$, which
depends on the choice of the reference dimer configuration.

\section{Discussion and conclusion}
\label{sec:disc-concl}

In this work, we study an exactly solvable model on the triangular
lattice with a gapped spin-liquid ground state that has the
double-semion topological order. Comparing to the original RK
Hamiltonian, this exactly solvable Hamiltonian contains the same local
resonance and potential terms, but the coefficient of the resonance
term becomes imaginary. As a result, the ground-state wave function of
this exactly solvable Hamiltonian has a nonlocal sign structure
determined by counting the number of loops in the transition
graph. Moreover, the wave function of monomer excitations also has a
sign structure that can be constructed using a half-vison
string. Using the sign structure in the ground state and the excited
states, we explicitly demonstrate that the monomer excitations obey
the fusion rules and the statistics of semion and antisemion
excitations. This implies that the ground state has the double-semion
topological order.

We constructed two forms of an exactly solvable Hamiltonian realizing the
same double-semion topological order, which are related by a local
unitary transformation. In one form, it explicitly breaks
translational and rotational symmetries of the lattice. The other form
retains the lattice symmetries but realizes time-reversal and mirror
symmetries in a twisted way.  Furthermore, we also constructed a
simple fully symmetric exactly solvable Hamiltonian on the star
lattice.

Our construction of dimer wave functions with double-semion
topological order can be generalized to construct spin wave functions
in order to study spin-liquid states. (We note that a spin wave
function with the double-semion topological order has been realized in
a spin-1 system using a different approach~\cite{Scharfenberger2011}.)
For antiferromagnetic spin-liquid states with a large spin gap, a spin
wave function analogous to the dimer wave function in
Eq.~\eqref{eq:rk-wf} can be constructed as a superposition of
different short-range spin-singlet pairing
patterns~\cite{Liang1988,Tang2011}. On a nonbipartite lattice, such wave
functions have the Z$_2$ topological order and a short-range spin-spin
correlation~\cite{Wildeboer2012,Yang2012}. The properties of such wave
functions can be studied numerically using the variational Monte
Carlo~\cite{Wildeboer2012,Yang2012} and projected entangled pair states
(PEPS)~\cite{Schuch2012,Wang2013} methods. A sign structure similar to the one
used in the dimer wave function in Eq.~\eqref{eq:wf} can be added to
this spin-liquid wave function and this additional sign structure will
change the topological order of the spin-liquid state to the
double-semion one. In particular, on kagome lattices, recent numerical
studies~\cite{HeShengChen2014,ShengSL,Bauer2014,He2014X} reveal
possible evidences of continuous phase transitions between the gapped
spin-liquid ground state of the nearest-neighbor Heisenberg model and
a time-reversal-symmetry-breaking chiral spin-liquid state, and it has
been proposed theoretically~\cite{Barkeshli2014} that such transitions
imply that the former spin liquid has the double-semion topological
order. Such a proposal can be investigated using the spin wave
function we just proposed, and we shall leave this interesting work to
future studies. (We note that since the double-semion topological
order necessarily breaks time-reversal or lattice
symmetries~\cite{Zaletel2015}, such a proposal implies that the gapped
spin-liquid ground state of the nearest-neighbor Heisenberg model
breaks time-reversal or lattice symmetries. This symmetry breaking is
not observed numerically and this issue needs to be resolved by future
DMRG studies.)

\emph{Note added}. After the completion of our work, we learned
that the realization of double-semion topological order on kagome lattice
quantum dimer models is also being considered by other
works~\cite{Buerschaper2014a, Iqbal2014}.

\begin{acknowledgments}
  We thank Liang Fu, Steve Kivelson, Chong Wang and Xiao-Gang Wen for
  invaluable discussions.  Y.Q. is supported by NSFC Grant
  No. 11104154. Z.-C. G. is supported by is supported by the
  Government of Canada through Industry Canada and by the Province of
  Ontario through the Ministry of Research and Innovation. H.Y. is
  supported in part by the National Thousand-Young-Talent Program.
\end{acknowledgments}

\appendix

\section{Symmetry of the double-semion ground state}
\label{sec:transl-invar}

In this appendix, we argue that the double-semion ground-state wave
function in Eq.~\eqref{eq:wf2} is invariant under the translation and
rotation symmetries of the triangular lattice.

Both the exactly solvable Hamiltonian in Eq.~\eqref{eq:eRK} and its
ground-state wave function in Eq.~\eqref{eq:wf} break the translation
and rotation symmetries of the triangular lattice because they depend
on the choice of a symmetry-breaking reference state, as explained in
Sec.~\ref{sec:sign-struct-dimer}. However, the translation and
rotation symmetries are restored in the Hamiltonian in
Eq.~\eqref{eq:iRK} after the local unitary transformation
$b_{ij}\rightarrow b_{ij}e^{i\theta_{ij}}$. After this local
transformation, the ground-state wave function takes the form in
Eq.~\eqref{eq:wf2}. Here we prove that this wave function is invariant
under lattice translation and rotations, although the form of
Eq.~\eqref{eq:wf2} does not manifest these symmetries explicitly.

To show this, consider that the exactly solvable Hamiltonian in
Eq.~\eqref{eq:iRK} is related to the original RK Hamiltonian by the
combination of the global unitary transformation in Eq.~\eqref{eq:ut}
and the aforementioned local unitary transformation. Therefore, the two
Hamiltonians have exactly the same spectrum. That is, all eigenstates
of the two Hamiltonians (including the ground state and excited states)
have the same energies and degeneracies, and the wave functions are
related by the combined unitary transformation. Now we consider the
ground states of the Hamiltonian. It is well known that for the
original RK Hamiltonian, if we neglect the non-flippable states, which
are disconnected with the rest of the states in the Hilbert space,
then there are four ground states, one in each topological section labeled
by the two winding numbers in the $x$ and $y$ directions. Hence there
is an unique nondegenerate ground state in each topological
sector. Because of the unitary transformation, the same claim also
holds for the Hamiltonian in Eq.~\eqref{eq:iRK}. Since the Hamiltonian
is itself translational and rotational invariant, after these symmetry
transformations the ground-state wave function must stay invariant, up
to a global phase factor.

Another way of arguing this is that since the Hamiltonian in
Eq.~\eqref{eq:eRK} respects translation and rotation symmetries, the
ground-state wave function can only break these symmetries through
spontaneous symmetry breaking, which would imply ground-state
degeneracy in the spectrum of the Hamiltonian. However, neither the
spectrum of this Hamiltonian nor that of the original RK Hamiltonian
have the ground-state degeneracy corresponding to the spontaneous
symmetry breaking. Note that the ground-state degeneracy due to
spontaneous symmetry breaking is not to be confused with the
topological ground-state degeneracy: the former degeneracy is between
states within the same topological sector defined by the winding
numbers, and the latter is between states in different topological
sectors.

\section{Half-vison string operator}
\label{sec:semion-string}

In this appendix, we discuss the definition and properties of the
string operators we use in Sec.~\ref{sec:monom-excit-stat} to
describe the sign structure of monomer wave functions. This string
operator is used to define the semion and antisemion excitations, and
it acts like the square root of a vison string. Therefore, we coin the
name ``half-vison string.'' In Sec.~\ref{sec:monom-excit-stat}, the
half-vison string operator is defined for the special cases that
contain only two monomers and the half-vison string operator connects
these two monomer sites. Here we give the definition for general cases
and argue that this definition reduces to the simplified version in
the special cases.

\begin{figure}[htbp]
  \centering
  \includegraphics{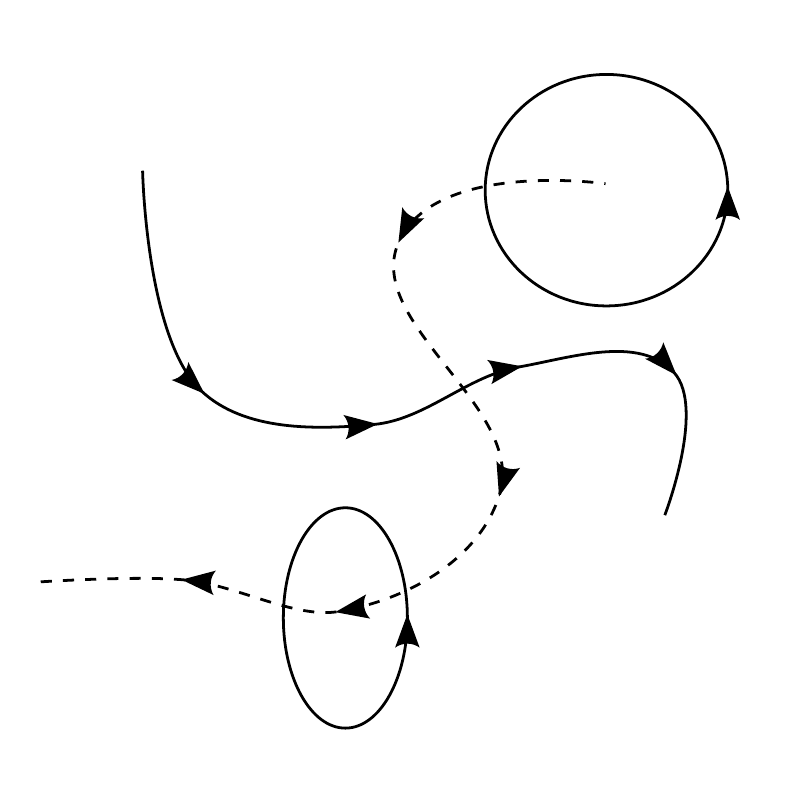}
  \caption{General definition of a half-vison string operator. The
    solid lines represent loops and monomer strings in the transition
    graph, and the dashed line represents a half-vison string.}
  \label{fig:hvstr}
\end{figure}

The general definition of the half-vison string operator is
illustrated in Fig.~\ref{fig:hvstr}. For any dimer configuration, we
construct the transition graph between it and the reference
configuration. In the transition graph, we assign orientations to both
the monomer strings and transition loops. In
Sec.~\ref{sec:monom-excit-stat}, we see that the orientation of a
monomer string determines the type of monomers at its ends. For the
transition loops, we always choose the orientation to be
counterclockwise. Then the result of the half-vison string operator is
the product of phase factors every time the half-vison string cuts a
dimer, and the phase factor depends on the angles between the
half-vison string and the dimer: the phase factor is $i$ if the dimer
is going from the left to the right relative to the direction of the
half-vison string, and the phase factor is $-i$ otherwise. From the
definition of the semion string operator, we can deduce the following
properties.

First, two semion string operators in the same direction becomes a
vison string, as the square of the phase factors $\pm i$ is always
$-1$. Hence a semion string can be viewed as the square root of a
vison string. We note that here we define a vison string as a string
operator that contributes a $-1$ sign every time it cuts a dimer in the
transition graph, where dimers from the reference configuration are
also counted. However, in most literature, it is defined without
counting the dimer in the reference state. Since the dimer
configuration in the reference state is fixed, the two definitions
only differ by an overall factor that does not depend on the dimer
configuration we consider.

Second, for dimer configurations with monomers at fixed locations, the
semion string operator depends only on the starting and ending of the
string and does not depend on the path. The difference between two
different semion strings starting and ending at the same locations is
a closed semion string loop. If such loop encloses no monomer sites,
the loop intersects any transition loop or string even times and the phase
factos cancel, so the closed string operator is equal to the identity operator, and
therefore the two semion string operators are equal. Moreover, one can
show that if the semion string loop encloses monomer excitations, the
string loop operator is equal to a constant phase factor, which is a
product from each enclosed monomer excitation, where a monomer with
an outgoing and incoming transition string contributes a $\pm i$,
respectively. Therefore, in this case, the two semion operators only
differ by an overall phase factor.

Lastly, the general definition of a half-vison string can be reduced
to the simplified form we use in Sec.~\ref{sec:monom-excit-stat},
where only intersections with the monomer string are considered, if
there are only two monomers and the half-vison string connects the two
monomer sites. Since the transition loops and the monomer strings do
not intersect, in this case there is no transition loop that encloses
any end of the half-vison string. Therefore, for every transition loop
that the half vison cuts, it cuts it even times and the total phase
factor cancels. Therefore, in this case only cuts, with the monomer
string needs to be considered. We note that when we say the half-vison
string starts or ends on a lattice site, there is an ambiguity on the
exact location of the starting or ending point because the half-vison
string lives on the dual lattice, and its starting (or ending) point
can only be on a dual lattice. Here one dual lattice point around the
monomer site can be arbitrarily chosen as the starting point of the
half-vison string. Since in the transition graph there is only one
dimer from the reference configuration connecting to the monomer site,
and the position of this reference dimer is therefore fixed, this
arbitrary choice of half-vison string starting (or ending) point does
not affect the definition of the half-vison string operator up to an
overall phase factor that does not depend on the dimer configuration.



\bibliography{double_semion,mybib}

\end{document}